\documentclass[aps,twocolumn,superscriptaddress,citeautoscript]{revtex4-1}
\usepackage{graphicx}
\usepackage{dcolumn}
\usepackage{amsmath}
\usepackage{amssymb}
\usepackage{wasysym}
\usepackage{color}

\usepackage{mathtools}

\begin{document} 

\title{Single reconstructed Fermi surface pocket in an underdoped single layer cuprate superconductor}

\author{M. K.~Chan}
\email{Correspondence to mkchan@lanl.gov and nharrison@lanl.gov}
\affiliation{Mail~Stop~E536,~Pulsed Field Facility, National High Magnetic Field Laboratory,~Los Alamos National Laboratory,~Los Alamos,~New Mexico 87545, USA}
\affiliation{School of Physics and Astronomy, University of Minnesota, Minneapolis, Minnesota 55455, USA}
\author{N. Harrison}
\email{Correspondence to mkchan@lanl.gov and nharrison@lanl.gov}
\affiliation{Mail~Stop~E536,~Pulsed Field Facility, National High Magnetic Field Laboratory,~Los Alamos National Laboratory,~Los Alamos,~New Mexico 87545, USA}\
\author{R. D. McDonald}
\affiliation{Mail~Stop~E536,~Pulsed Field Facility, National High Magnetic Field Laboratory,~Los Alamos National Laboratory,~Los Alamos,~New Mexico 87545, USA}
\author{B. J. Ramshaw}
\affiliation{Mail~Stop~E536,~Pulsed Field Facility, National High Magnetic Field Laboratory,~Los Alamos National Laboratory,~Los Alamos,~New Mexico 87545, USA}
\author{K. A. Modic}
\affiliation{Mail~Stop~E536,~Pulsed Field Facility, National High Magnetic Field Laboratory,~Los Alamos National Laboratory,~Los Alamos,~New Mexico 87545, USA}
\author{N. Bari{\v s}i{\'c}}
\affiliation{Technishce Universit\"{a}t Wein, Wiedner Haupstr. 8-10, 1040, Vienna, Austria}
\affiliation{School of Physics and Astronomy, University of Minnesota, Minneapolis, Minnesota 55455, USA}
\author{M. Greven}
\affiliation{School of Physics and Astronomy, University of Minnesota, Minneapolis, Minnesota 55455, USA}

\begin{abstract}
The observation of a reconstructed Fermi surface {\it via} quantum oscillations in hole-doped cuprates opened a path towards identifying broken symmetry states in the pseudogap regime. However, such an identification has remained inconclusive due to the multi-frequency quantum oscillation spectra and complications accounting for bilayer effects in most studies. We overcome these impediments with high resolution measurements on the structurally simpler cuprate HgBa$_2$CuO$_{4+\delta}$ (Hg1201), which features one CuO$_2$ plane per unit cell. We find only a single oscillatory component with no signatures of magnetic breakdown tunneling to additional orbits. Therefore, the Fermi surface comprises a single quasi-two-dimensional pocket. Quantitative modeling of these results indicates that biaxial charge-density-wave within each CuO$_2$ plane is responsible for the reconstruction, and rules out criss-crossed charge stripes between layers as a viable alternative in Hg1201. Lastly, we determine that the characteristic gap between reconstructed pockets is a significant fraction of the pseudogap energy.

\end{abstract}

\maketitle

\section{Introduction}
The identification of broken symmetry states, particularly in the pseudogap region, is essential for understanding the cuprate phase diagram. The surprising discovery of a small Fermi surface from quantum oscillations (QOs) in underdoped YBa$_2$Cu$_3$O$_{6+x}$ (Y123)~\cite{leyraud07} motivated proposals for a crystal-lattice-symmetry-breaking order parameter~\cite{millis07,chakravarty08,galitski09,yao11,podolsky08,harrison09,sebastian14,allais14} that reconstructs either the large Fermi surface identified in overdoped cuprates~\cite{hussey03,vignolle08,hossain08} or the Fermi arcs of the pseudogap state~\cite{ding96,loeser96,punk15}.

The spectrum of quantum oscillations is, in principle, a distinct probe of the Fermi surface morphology and is thus a signature of the broken symmetry state~\cite{millis07,chakravarty08,millis07,chakravarty08,galitski09,yao11,sebastian14,allais14,podolsky08,harrison09}. The ubiquity of short-range charge-density-wave (CDW) in underdoped cuprates~\cite{ghiringhelli12,chang12,tabis14,campi15,comin14,dasilva14,hoffman02} make CDW a natural choice as the order responsible for Fermi surface reconstruction. However, despite the availability of exquisitely detailed QO studies in Y123~\cite{audouard09,ramshaw11,sebastian14,doiron15}, its complicated multi-frequency spectrum has prevented a consensus on the exact model for reconstruction~\cite{ramshaw11,sebastian14,harrison15,maharaj15,briffa15,allais14,doiron15}. Part of the difficulty stems from the crystal structure of Y123, particularly the bilayer splitting of the elementary Fermi pockets due to the two CuO$_2$ planes per unit cell. The different models are sensitive to the magnitude, symmetry, and momentum dependence of the bilayer coupling~\cite{sebastian14,harrison15,briffa15,maharaj15}, which are controversial. Furthermore, neither diffraction nor QO experiments in the cuprates have yet been able to address the crucial question as to whether the two orthogonal CDW vectors spatially coexist  in the same CuO$_2$ plane or whether stripes alternate in a criss-cross fashion on consecutive CuO$_2$ planes~\cite{hosur13,maharaj14}. 

Apart from the Y-based bilayer compounds~\cite{leyraud07,yelland08}, HgBa$_2$CuO$_{4+\delta}$ (Hg1201) is the only other hole-doped cuprate for which QOs have been detected~\cite{barisic13b}  in the pseudogap regime. Importantly, in addition to featuring a  very high-$T_{\rm c}$ ($\approx 97$~K at optimal doping),  Hg1201 has a tetragonal crystal symmetry consisting of only one CuO$_2$ plane per unit cell. This means that the analysis of the experimental data on this compound is free from complications associated with bilayer coupling and orthorombicity.

Here we show that high resolution measurements of up to 10 cycles of the QOs in Hg1201 permit a resolution of the reconstructed electronic structure. Using pulsed magnetic fields extending to 90~T combined with contactless resistivity measurements, we find the QOs in Hg1201 to be remarkably simple: a single oscillation frequency exhibiting a monotonic magnetic field dependence characteristic of a single Fermi surface pocket. We find quantitative agreement between the observed single QO frequency and that from a diamond-shaped electron pocket resulting from biaxial CDW reconstruction~\cite{harrison11,tabis14}. There are no signatures of the predicted additional small hole-like pocket~\cite{allais14} reported for Y123~\cite{doiron15}. This could be due to the antinodal states, which constitute these hole pockets, being gapped out or strongly supressed by the pseudogap phenomena. We also determine a very small $c$-axis transfer integral for Hg1201, which precludes a model based on an alternating criss-cross pattern of uniaxial charge stripes on consecutive CuO$_2$ planes~\cite{maharaj14}. The absence of signatures of magnetic breakdown tunneling to neighboring sections of the Fermi surfaces (such as the putative small hole pockets~\cite{allais14}) provides a lower bound estimate of $\approx$~20~meV for the relevant gap. Importantly, this is  a significant fraction of the anti-nodal pseudogap energy~\cite{timusk99}. Overall, our results point to biaxial CDW reconstruction acting on the short nodal Fermi arcs produced by the pseudogap phenomena.

\section{Results}
\begin{figure}[ht!!!!!!!!!!] 
\includegraphics{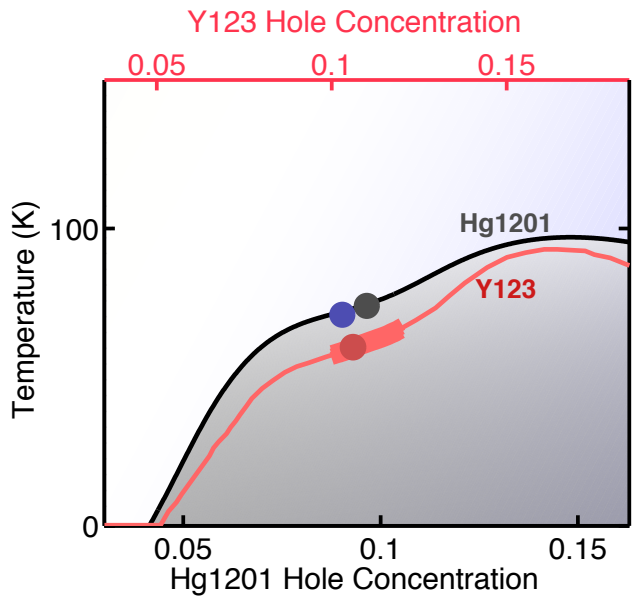}%
\caption{
{\bf Hole concentration of Hg1201 and Y123 samples.} Superconducting temperature , $T_{\rm c} (p)$, as a function of hole concentration $p$ for Hg1201~\cite{yamamoto00} (black line, bottom x-axes) and Y123~\cite{liang06} (red line, top x-axes). $T_{\rm c}(p)$ of  Hg1201 UD71 and UD74 highlighted in the current study of the topology of Fermi-surface reconstruction are indicated by blue and black circles respectively. The doping of the Y123 sample studied here is indicated by the red circle. Although quantum oscillations for Y123 have been reported over a wider range of hole concentrations~\cite{sebastian10c,ramshaw15}, detailed studies of the spectra have focused on the narrow range indicated by the thick red line~\cite{sebastian08,audouard09,ramshaw11,sebastian14,doiron15}, corresponding to the plateau on the  $T_{\rm c}(p)$ dome and where the amplitude of oscillations is largest. \label{pd}} 
\end{figure}

\begin{figure}[ht!!!!!!!!!!] 
\centering
\includegraphics*{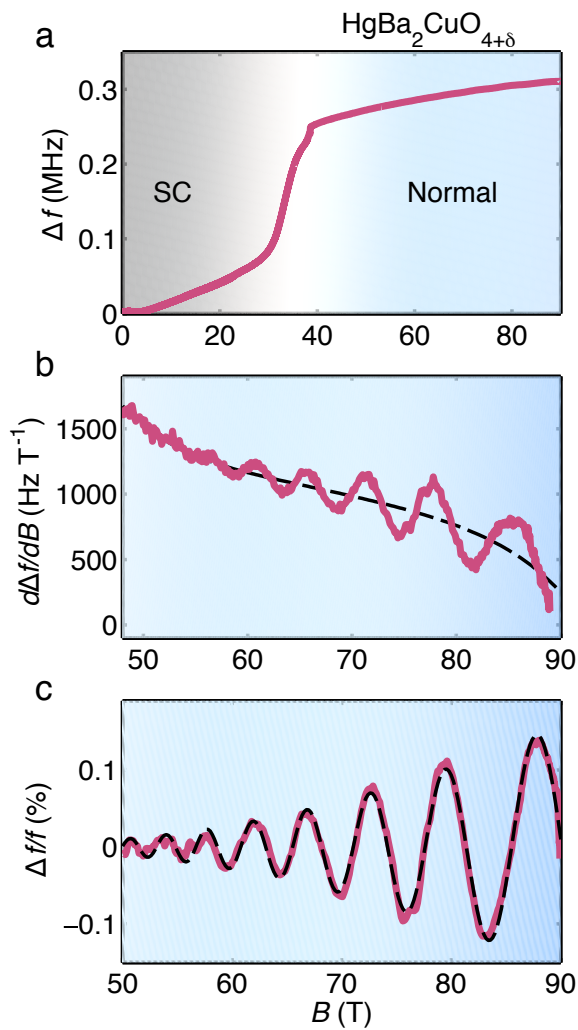}%
\caption{
{\bf Observation of quantum oscillations in Hg1201 with contactless resistivity.} (a) Evolution of the PDO circuit frequency coupled to Hg1201 UD71 with applied magnetic field $B$ along the c axis of the sample at $T = 1.8$ K. The sample undergoes a transition from superconducting (SC, black shaded region) to normal (blue region) at $B\sim 35$~T. (b) Derivative of the raw data with respect to magnetic field reveals quantum oscillations in the normal state. As described in the text, a non-oscillatory polynomial background is subtracted from the raw data to extract the quantum oscillations. The derivative of the background is shown as the dashed black line. (c) Quantum oscillations after the polynomial background has been removed. The dashed black line is a fit to the Lifshitz-Kosevitch form discussed in Methods.}
\label{raw}
\end{figure}

{\bf Quantum oscillation measurements in pulsed magnetic fields.} 
The typical sample quality and magnetic field requirements for observing QOs is exponentially dependent on the condition $\omega_c\tau\gtrsim1$, where $\omega_c = e B/m^\star$ is the cyclotron frequency and $1/\tau$ is the scattering rate. For the Hg1201 samples studied here, $\omega_c\tau\approx 0.35$ at $B=45~{\rm T}$ on average. Thus, compared to Y123, which has $\omega_c\tau\approx 1.2$~\cite{sebastian11} at the same field, QO measurements in Hg1201 are much more challenging. To overcome this, we utilize a high sensitivity contactless resistivity method wherein the sample forms part of a proximity detector oscillator (PDO) circuit~\cite{altarawneh09}, and extremely large magnetic fields. Changes $\Delta f$ in the PDO circuit frequency $f$ in an applied magnetic field $B$ are directly related to the changes in the complex penetration depth of the sample~\cite{coffey92} and hence the in-plane resistivity. We focus on hole-doping where a plateau in the $T_{\rm c}$ dome occurs , which is also the region where detailed QO measurements in Y123 have focused, as indicated in Fig.~\ref{pd}. Fig.~\ref{raw}(a) shows $\Delta f/f$ for an underdoped Hg1201 sample UD71 ($T_{\rm c}=71$~K) in an applied magnetic field. The large increase in $\Delta f/f$ at $B\approx 35$~T corresponds to the transition from the superconducting to the resistive state. A derivative of the data with respect to magnetic field clearly reveals QOs in the resistive state, without the need for background removal (see Fig.~\ref{raw}(b)). 

\begin{figure*}[ht!!!!!!!!!] 
\includegraphics[width=1\textwidth]{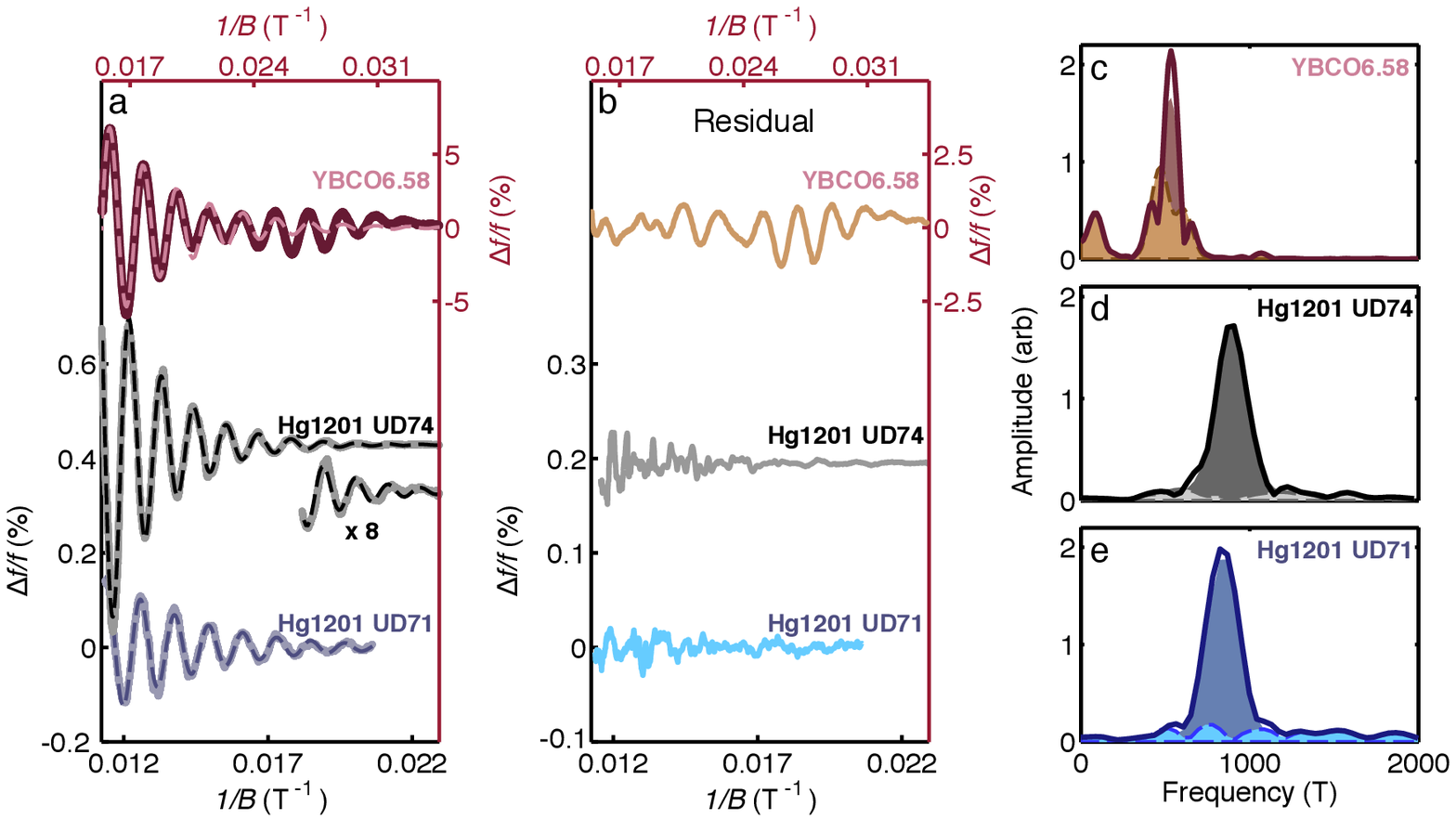}%
\caption{
{\bf Spectrum of quantum oscillations in Hg1201 and Y123}(a) Percentage change of the resonance frequency  $\Delta f/f$ as a function of inverse applied field $1/B$ for PDO circuits coupled to Y123 with $x=$~0.58 ($T_{\rm c} = 60$~K, red), Hg1201 UD74 ($T_{\rm c} = 74~$K; black) and Hg1201 UD71 ($T_{\rm c} = 71~$K; blue). The small amplitude oscillations for Hg1201 UD74 at low fields (large $1/B$) are magnified $\times8$ for clarity. Single frequency fits to the data with Eq.(1) (see Methods) are indicated by dashed lines. (b) The residual after subtracting the single frequency fit from the data in (a). Note the smaller vertical scale of the axis. (c)-(e) Fourier transform for Y123 with $x=$~0.58, Hg1201 UD74, and Hg1201 UD71. The solid lines are FFT of the data and  the dark shaded regions represent the FFT of the single-frequency fits in (a). The light shaded regions are the FFT of the residual in (b). \label{osc}}
\end{figure*}

{\bf Single quantum oscillation frequency in Hg1201.} 
Figure \ref{raw}(c) shows QOs after removing the background as described in Methods. The dashed line is a good fit of the data (solid line) to the expected QO waveform for a single quasi-two-dimensional Fermi surface with no warping and no magnetic breakdown tunneling (see Eq~.\ref{lk1} in Methods). Warping and magnetic breakdown introduces other frequency components manifest as a beat or non-monotonic amplitude modulation, which are absent in our data. In Fig.~\ref{osc}a, we show high resolution data obtained by averaging multiple magnetic field shots for UD71 and an additional Hg1201 sample of slightly higher $T_{\rm c}$ ($T_{\rm c} = 74$~K, labeled UD74). Seven and ten full oscillations are resolved for UD71 and UD74, respectively. For both samples, the observed QOs are well captured by the fit (dotted lines in Fig.~\ref{osc}(a)), which yields oscillation frequencies of  $F=847(15)$~T and $893(15)$~T for UD71 and UD74. 

In order to set limits on the amplitude of additional QO frequencies, we determine the residuals by subtracting the single frequency fits from the data in Fig.~\ref{osc}b. The residuals for both Hg1201 samples do not show evidence for additional oscillatory components. On further comparing the Fourier transform of the data with the Fourier transform of the fit (dark shaded regions in Fig.~\ref{osc}d and \ref{osc}e), both can be seen to have the same line shape, thereby providing further evidence for the absence of additional quantum oscillation frequencies. The Fourier transform of the residuals (light shaded regions in Fig.~\ref{osc}d and \ref{osc}e) are devoid of prominent peaks, consistent with it representing the noise floor of the experiment. From the noise floor, we can infer that in order for additional quantum oscillation frequencies to go undetected, they must fall below $\approx 8~\%$ of the Fourier amplitude of the observed QO frequency. It is tempting to attribute the weak modulating background we observe in the derivative of the raw data in sample UD71 (Fig.\ref{raw}(b)), to the additional low frequency oscillation previously reported for Y123~\cite{doiron13}. However, as discussed in Supplementary Note 1 and Supplementary Figures 1\&2 , the lack of such a modulation in sample UD74 and the disappearance of the modulation in UD71 at slightly elevated temperatures (in contradiction with the light mass of the small pocket reported for Y123~\cite{doiron13}), leads us to conclude that this is a background effect unrelated to slow QOs.

Our finding of a single frequency in Hg1201 contrasts significantly with the multiple frequencies present in Y123~\cite{audouard09,sebastian12} evident from the data with the same number of oscillation periods as for Hg1201 UD74 (see Figs.~\ref{osc}). In contrast to the residuals obtained for Hg1201, the residual for Y123 (see Fig.~\ref{osc}b), again obtained on subtracting a fit to the dominant quantum oscillation frequency (of $F\approx$~530~T), reveals a distinctive beat pattern resulting from the interference between the two remaining QO components whose amplitudes are $\approx$~40 and 50~\% of the dominant frequency (FT of the residual shown Fig.~\ref{osc}c).

\begin{figure}[t!!!!!!!!!!] 
\includegraphics{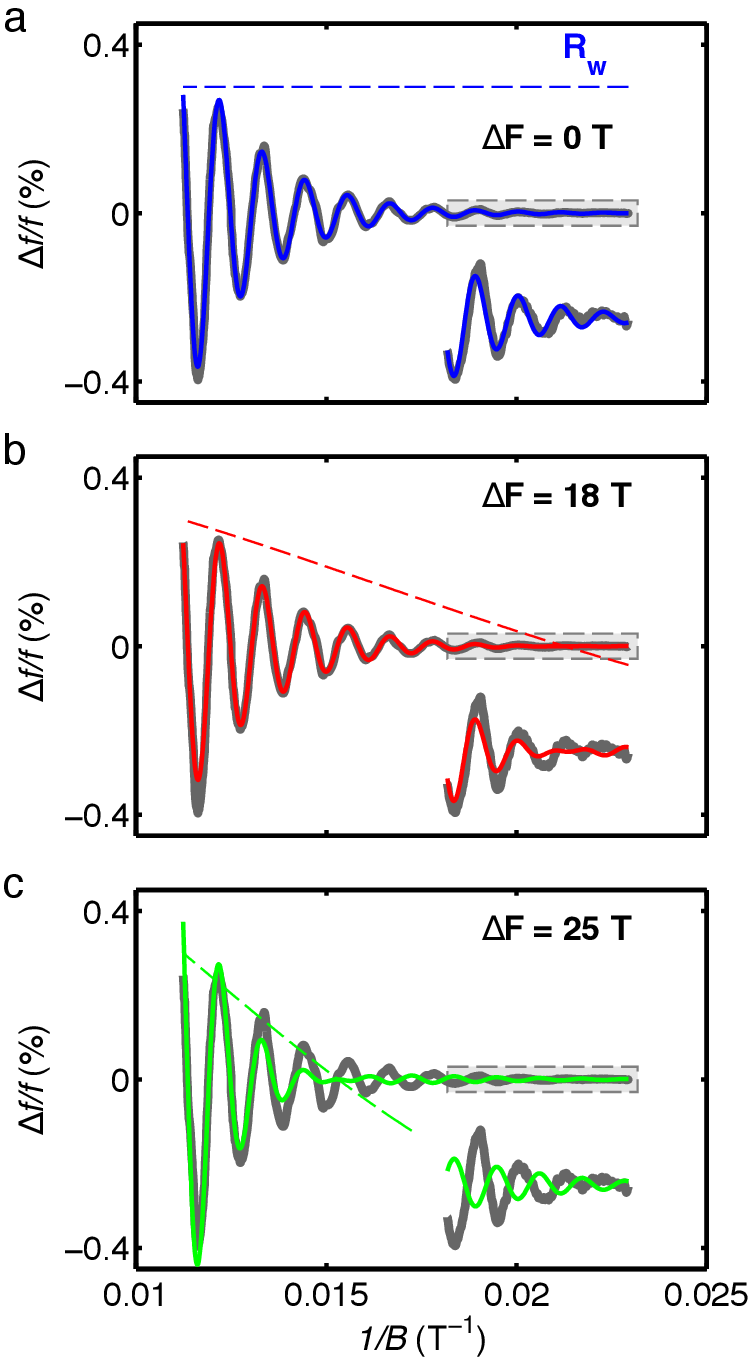}%
\caption{
{\bf Effect of warping on the quantum oscillations.} Influence of 2D Fermi surface warping on fitting QOs in Hg1201 (a) Fit (blue line) of the QO data (grey line) for UD74 with no warping ($\Delta F_c = 0$). The dotted line represents the $1/B$ dependence in arbitary units of the warping term $ R_{\rm w} = J_0 (2\pi\frac{\Delta F_c}{B})$ (see Methods). The degree of warping is increased for (b) and (c) yielding nodes in the QO spectra corresponding to $1/B$ values where $ R_{\rm w}$ changes sign . The small amplitude oscillations at large $1/B$, bordered by the grey box, is magnified by a factor of 15 in all three panels and shown as insets. \label{Supwarp}}
\end{figure}

\begin{figure}[t!!!!!!!!!!] 
\includegraphics{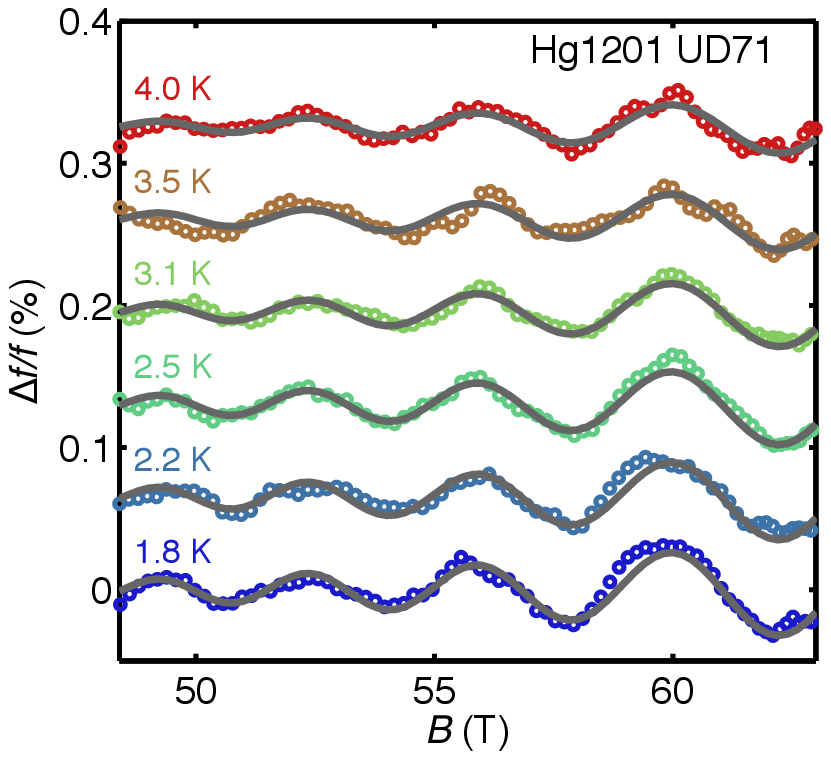}%
\caption{
{\bf Determination of the effective mass.} Relative PDO circuit frequency change $\Delta f/f$ as a function of field $B$  for sample Hg1201 UD71 at the temperatures indicated above each curve. Solid lines are simultaneous fits with Eq.(1) (see Methods) to the total data set where all parameters are constrained to be temperature independent. For the single frequency spectrum of Hg1201, this method produces a more robust determination of the effective mass than examining the temperature dependence of the FFT amplitude. The effective mass is extracted do be $m^\star/m_e = 2.7\pm0.1$, where $m_e$ is the bare electron mass. This is slightly larger than that determined for a sample with almost the same doping~\cite{barisic13b}. \label{Tdep}}
\end{figure}

{\bf Limits on the Fermi surface warping and $c$-axis hopping}. 
In quasi-two-dimensional metals, the inter-plane hopping leads to warping of the cylindrical Fermi surface, yielding two oscillation frequencies originating from minimum and maximum extremal cross-sections. While our observation of a single quantum oscillation frequency rules out very large warping, small warping can manifest in observable nodes in the  magnetic field-dependent QO amplitude.  This is represented by an additional amplitude factor, $R_{\rm w}$, which is parametrized by the separation between the two frequencies $2\Delta F_{\rm c}$ (see Methods). To illustrate this point, in Fig.~\ref{Supwarp} we fit the data with several different fixed values of $\Delta F_{\rm c}$ in $R_{\rm w}$. The absence of nodes in the experimental data enables us to make an upper bound estimate of $\Delta F_{\rm c}<$~16~T.
Using $m^*\approx 2.7~m_{\rm e}$ (see Fig.\ref{Tdep}) and $2\Delta F_{\rm c}\approx 4t_\perp m^\ast/({\hbar e})$, we obtain a $c$-axis hopping of $t_{\perp}<0.35$~meV for Hg1201, revealing it to be at least 1000 times smaller than the nearest neighbor hopping ($t = 460$~meV~\cite{das12}) within the CuO$_2$ planes. Our upper bound is also 25 times smaller than the bare value determined from LDA calculations ($t_\perp = 10$~meV~\cite{das12}), reflecting a large quasiparticle re-normalization.

Our ability to set a firm upper bound estimate for $t_{\perp}$ in Hg1201 contrasts with the situation in 
Y123, where estimates of the $c$-axis warping are challenging to separate from the effects of bilayer coupling. Estimates for $\Delta F_{\rm c}$ in Y123 range from $\approx$~15 to 90~T depending on whether the observed beat pattern originates from the combined effects of bilayer-splitting and magnetic breakdown tunneling~\cite{sebastian14} or Fermi surface warping~\cite{audouard09,ramshaw11,doiron15}. 

\begin{figure*}[t!!!!!!!!!!] 
\includegraphics*{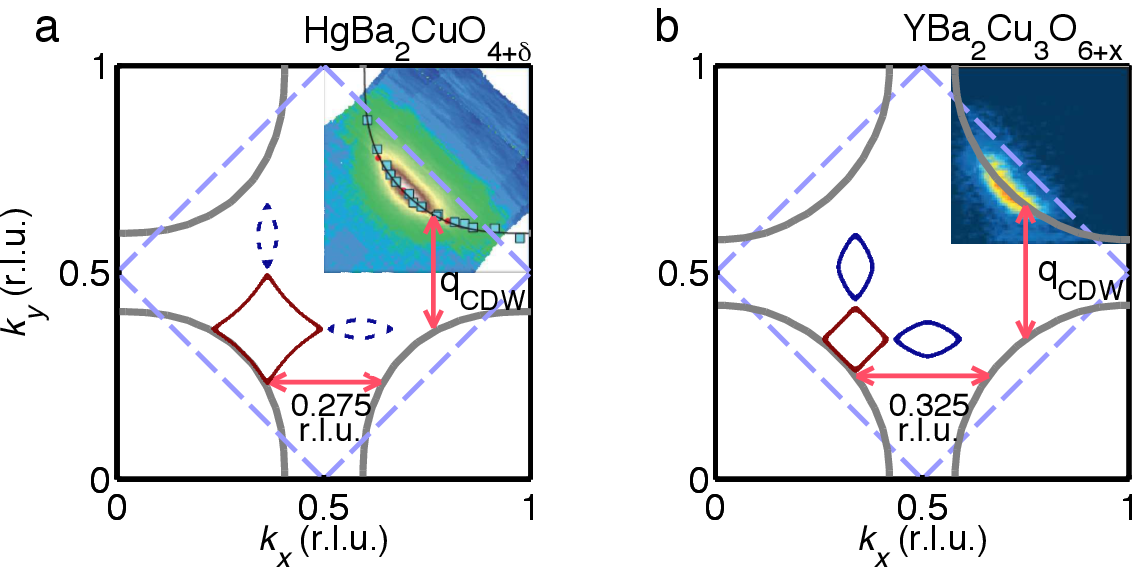}
\caption{{\bf Fermi surface reconstruction by charge-density-wave} Unreconstructed (grey lines) and reconstructed (colored lines) Fermi surface of HgBa$_2$CuO$_{4+\delta}$ ($p=0.095$) (a) and YBa$_2$Cu$_3$O$_{6+x}$ ($p=0.11$) (b). The single-frequency QO we observe for Hg1201 is in agreement with the area of the electron-like diamond shaped pocket (solid red), while there are no signatures of the small hole-like pocket (dashed dark blue) in our data. This might be due to the lack of quasiparticle weight in the pseudogapped antinodal regions of the Fermi-surface determined from angle-resolved photoemission~\cite{ding96,loeser96,punk15}. For the reconstruction, we assume biaxial CDW wavevectors $(Q_{\rm CDW},0)$ and $(0,Q_{\rm CDW})$. $Q_{\rm CDW}=0.275$ r.l.u. for  Hg1201 ($T_{\rm c} = 72$~K) was taken from Ref.\cite{tabis14}. $Q_{\rm CDW}=0.325$ for Y123 is estimated based on measurements on Y123 with $x=$~0.54\cite{blackburn13} and  Y123 with $x=$~0.55 \cite{blanco14}. The color plots in the upper right edges of the panels are photoemission data showing the Fermi surface map for Hg1201 ($p\approx0.12$) (adapted with permission from Ref. \cite{vishik14}. Copyrighted by the American Physical Society) and Y123  ($p=0.11$) (adapted by permission from Macmillan Publishers Ltd: Nature (Ref.~\cite{hossain08}), copyright (2008)). The points on top of the angle-resolved photoemission data in (a) are Fermi surface crossings. The tight-binding hopping parameters were determined by fitting the photoemission data while constraining the area of the Fermi surface to match the quoted hole concentrations.  Dashed blue lines indicate the antiferromagnetic  AF zone boundaries.}
\label{schematic}
\end{figure*}

{\bf Fermi surface reconstruction by biaxial CDW.} 
The simple crystalline structure of Hg1201 makes it the ideal system for relating the $k$-space area of the observed Fermi surface pocket to prior photoemission~\cite{vishik14} and x-ray scattering~\cite{tabis14} measurements. The former constrains the unreconstructed Fermi-surface while the later provides the magnitude of the reconstruction wave-vector. Following Allais {\it et al.}~\cite{allais14}, we require that the large unreconstructed hole-like Fermi surface of area $A_{\rm UFS}$ accommodate $1+p$ carriers, where $p$ is the hole doping defined relative to the half filled band. We then proceed to translate the Fermi surface multiple times by the wavevectors $(Q_{\rm CDW},0)$ and $(0,Q_{\rm CDW})$ and their combinations in Fig.~\ref{schematic}a. Here we have assumed a biaxial reconstruction scheme. Further details of the calculation is described in the  Methods section.

The biaxial CDW reconstruction (shown in Fig.~\ref{schematic}) yields a diamond-shaped electron pocket (depicted in red) flanked by smaller hole pockets (depicted in blue) accompanied by additional open Fermi surface sheets (shown in Supplementary Figure 3). While the parameters for Hg1201 are slightly different than for Y123, the topology of the reconstructed Fermi surface is essentially the same.

Using the Onsager relation $F_{\rm e,h} = \frac{\hbar}{2\pi e}A_{\rm e,h}$, where $A_{\rm e}$ is the area of the  electron pocket and $A_{\rm h}$ is the area of the hole pocket, the calculated QO frequencies are $F_{\rm e} =885$~T and $F_{\rm h} =82$~T for the electron and hole pocket. $F_e$ is remarkably close to that observed for Hg1201: $847$~T and $893$~T for UD71 and UD74, respectively. However, $F_h$ is not observed in our experiment (see Supplementary Note 1 and Supplementary Figures~1\&2). 

If we instead consider a purely uniaxial reconstruction, i.e. stripe CDW order, our calculation with the same parameters for Hg1201 yields an oval-shaped hole-like pocket at the anti-nodal regions of the original Fermi surface with a frequency of 590 T and no futher closed pockets (see Supplemental Figure 4) . This is in much poorer agreement to the measured QO frequency, and furthermore, disagrees with Hall effect measurements which imply the existence of a predominantly electron-like reconstructed Fermi surface~\cite{doiron13}.

In Fig.~\ref{schematic}(b) we show the same calculation for Y123 with $x=$~0.58 ($p \approx 0.106$). The agreement between the measured dominant frequency ($534$ T) and calculated electron-pocket frequency $F_{\rm e} =380$~T is less satisfactory. However, here (like Allais~{\it el al.}~\cite{allais14}) we have neglected some of the complications in Y123 such as the bilayer-coupling and orthorhombic crystal structure. The photoemission data are also made more difficult to interpret in Y123 owing to the necessity of surface K doping to reach the desired hole doping of $p\approx$~11~\%~\cite{hossain08}. These uncertainties highlight the utility of studying the structurally simpler Hg1201.

{\bf Limits on magnetic breakdown tunneling across band gaps.} 
It was recently proposed by Allais {\it et al.}~\cite{allais14} that a sufficiently large magnetic field enables magnetic breakdown tunneling to occur between the electron and hole pockets shown in Fig.~\ref{schematic}, providing a possible explanation for one or more of the observed cluster of three frequencies in Y123~\cite{doiron15}. The probability of magnetic breakdown tunneling increases with magnetic field, giving rise to new orbits and a reduction in the elementary QO amplitudes by $R_{\rm MB} = (i\sqrt{P})^{n_p} (\sqrt{1-P})^{n_q}$. Here,  $i\sqrt{P}$ and $\sqrt{1-P}$ are the tunneling and reflection amplitudes respectively, while $n_p$ and $n_q$ are the number of breakdown tunneling and Bragg reflection events encountered {\it en route} around the orbit~\cite{shoenberg}. The magnetic breakdown probability is given by $P =\exp\{-B_0/B\}$, where $B_0 = ({m^\ast}/{\hbar e })({E_{\rm g}^2}/{E_{\rm F}})$ is the characteristic breakdown field, in which $E_{\rm F}\approx\hbar eF_{\rm e}/m^\ast$ is the approximate Fermi energy and $E_{\rm g}$ is the band gap separating adjacent sections of Fermi surface. 

\begin{figure}[t!!!!!!!!!!] 
\includegraphics{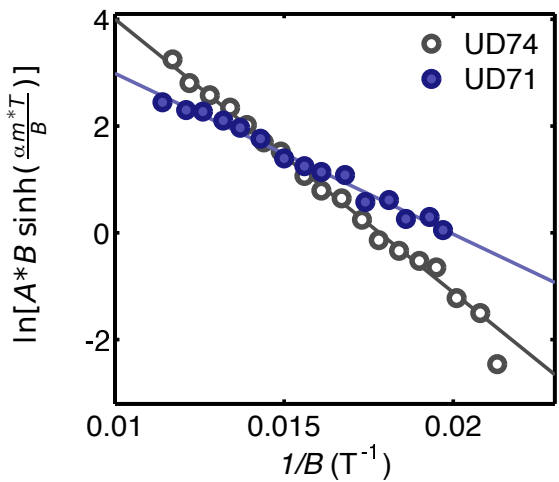}%
\caption{
{\bf Dingle plot.}Dingle plot of the QO amplitude normalized by $R_T$ as a function of $1/B$ for UD71 and UD74. Amplitude of maxima and minima are taken from the data in Fig.~\ref{osc}a. Solid lines are fits to the data as described in the text. \label{LL}}
\end{figure}

Magnetic breakdown can manifest itself in two ways in our data. The first, is by way of a reduction of the primary QO amplitude at higher magnetic fields.  For the diamond-shaped electron pocket  in Fig.~\ref{schematic}, the amplitude is reduced by $R_{\rm MB} = [\sqrt{1-{\rm exp}\{-B_0/B\}}]^4$ to account for the four Bragg-reflection points at the tips of the diamond. In a Dingle plot of the magnetic field dependence of the QO amplitude (Fig.~\ref{LL}), this should be discerned as deviations from a straight line for small $1/B$. Accordingly, we fit the Dingle plot to $a-b(1/B)+2{\rm ln}[1-{\exp}\{-B_0/B\}]$ where $b= \pi(\sqrt{2 \hbar F/e})/l$ (solid lines in Fig.~\ref{LL}b), yielding $B_0\approx 200$~T and $\approx$~250~T for UD74 and UD71, respectively. These large values for $B_0$ are consistent with no observable effects of magnetic breakdown (i.e. a straight-line Dingle plot), therefore, we take the fitted $B_0$ as lower bound values.

The second manifestation of magnetic breakdown is through the appearance of new QO frequencies corresponding to sums and differences of the areas of the Fermi surfaces involved in the tunneling process. Magnetic breakdown between the electron and hole pocket in our reconstruction model~\cite{allais14,doiron15}, results in additional QO frequencies of the form $F_{{\rm e-h},n} = F_{\rm e}-nF_{\rm h}$, in which $n$ is an integer. Based on our modeling, $F_{\rm h}\approx 80$~T, meaning that the frequencies $F_{{\rm e-h},n}$ are sufficiently distinct from $F_{\rm e}$ to be discernible in the raw and Fourier transformed data in Fig.~\ref{osc}b,d,e. The noise floor of $\approx$~8~\% of the dominant oscillation amplitude $A_{\rm e}$ in Fig.~\ref{osc} provides an upper limit for the amplitude $A_{{\rm e-h},1}$ the leading magnetic breakdown frequency ($n=1$). Using the inequality
 \[\frac{A_{e-h}}{A_e} =\frac{2 R_{\rm MB}^{e-h}}{R_{\rm MB}^{e}}=\frac{2 (i\sqrt{P})^{2} (\sqrt{1-P})^{4}}{(\sqrt{1-P})^{4}} \lesssim 0.08,\] we obtain a second lower bound of $B_0\approx$~200~T. Here, we have assumed a similar $m^\ast$ and scattering rate values for the various combination orbits, while the factor of two for $A_{\rm e-h}$ accounts for the two possible orbits involving one of the two hole-like pockets in Fig.~\ref{schematic}a.  
 
We have shown above that both the Dingle plot and the absence of additional Fourier peaks above the noise floor provide mutually consistent large lower bound estimates for $B_0$. The most conservative of these (i.e. $B_0\gtrsim$~200~T) enables a lower bound estimate of $E_{\rm g}\gtrsim$~20~meV to be made for the band gap between the observed electron and presumed hole pockets in Fig.~\ref{schematic}.

\section{Discussion}
Our observation of a simple monotonic waveform of a single  QO frequency in Hg1201, and a single Fermi surface cross-sectional area that is compatible with photoemission and X-ray scattering measurements, is essential for resolving issues relating to the nature of the CDW ordering. One of these concerns whether the two charge ordering wavevectors $(Q_{\rm CDW},0)$ and $(0,Q_{\rm CDW})$ coexist in the same CuO$_2$ plane~\cite{harrison11,allais14, briffa15} or whether stripes alternate in a criss-cross fashion on consecutive CuO$_2$ planes~\cite{maharaj14,hosur13}. In the absence of a coupling between CuO$_2$ planes, criss-cross stripes lead to open Fermi surface sheets running in orthogonal directions on adjacent planes. The effect of the inter-plane coupling is to introduce a hybridization~\cite{maharaj14}. Whereas a strong coupling in the range $\sim$~10 to 100~meV occurs within the bilayers in Y123 and Y124~\cite{andersen95}, no such coupling occurs in single layer Hg1201 and only a very weak coupling provided by the interlayer $c$-axis hopping determined here to be $t_{\rm c}<$~0.4~meV can exist. In the context of criss-cross stripe order, the effect of such a weak $c$-axis hopping is to introduce a very small gap of order $t_{\rm c}$ in magnitude between the electron and hole pockets in Fig.~\ref{schematic},  which would then have a very small characteristic magnetic breakdown field of $B_0\sim$~0.1~T (several orders of magnitude smaller than the lower bound constraint on $B_0$ determined from our experimental results).  The magnetic breakdown amplitude reduction factor $R_{\rm MB}$ for $B_0=$~0.1~T would be so small that it would render the electron pocket not observable in experimentally relevant magnetic fields of $\gtrsim$~40~T. Our observations of a single electron pocket and small $c$-axis hopping therefore rule out criss-cross stripes as a viable route for creating observable Fermi surface pockets in Hg1201 at high magnetic fields.

Although the CDW correlations detected with X-ray scattering are a natural candidate for the cause of Fermi-surface reconstruction, an open question concerns if the correlation length is sufficiently large to support quantum oscillations. A small correlation length of the order parameter can manifest as additional damping of the QO amplitude in the Dingle term (see Methods), thus suppressing the effective mean free path $l$~\cite{harrison07}. For Y123, the CDW correlation length at $T_{\rm c}$ and $B=0~\rm{T}$ is $\xi_{\rm{CDW}}\approx65\rm~{\AA}$~\cite{ghiringhelli12}.The effective mean free path, $l\approx 200{~\rm \AA}$~\cite{sebastian11} obtained from QO measurements is of the same order of magnitude as $\xi_{\rm{CDW}}$. For Hg1201 both $\xi_{\rm{CDW}}$ and $l$ are similarly reduced compared to Y123:  $\xi_{\rm{CDW}}\approx20\rm~{\AA}$~\cite{tabis14} at $T\sim T_{\rm c}$ and $l = 85~\rm{\AA}$ (average of UD71 and UD74 and consistent with prior Hg1201 results~\cite{barisic13b}). Thus, it appears that the QO effective mean free path might be correlated with the CDW domain size. Alternatively, both $l$ and $\xi_{\rm CDW}$ could be similarly affected by disorder or impurities. The relatively small $l$ for Hg1201 indicates that the CDW need not be long-ranged, even at low temperatures and high magnetic fields, to yield the reconstructed Fermi surface observed here. Although $\xi_{\rm{CDW}}$ in Y123 increases at low temperatures and high magnetic fields, it remains rather small ($\approx100~\rm{\AA}-400~\rm{\AA}$)~\cite{gerber15,chang16}.

Another issue concerns the origin of the $E_{\rm g}\gtrsim$~20~meV gap separating the diamond-shaped electron pocket from adjacent sections of Fermi surface in Hg1201. There are two possible CDW Fermi surface reconstruction scenarios that have been discussed in the literature. One of these involves the folding of the large Fermi surface~\cite{allais14}, as shown in Fig.~\ref{schematic}, which is expected to produce small hole pockets and open sheets in addition to the observed electron pocket. In such a scenario, $E_{\rm g}$ would then simply correspond to the CDW gap $2\Delta_{\rm CDW}$. The alternative scenario is that the reconstructed Fermi surface occurs by connecting the tips of Fermi arcs produced by a pre-existing or coexisting pseudogap state~\cite{comin14,tabis14}. In this scenario, we would expect the small hole pockets to be gaped out by the pseudogap causing $E_{\rm g}$ then to correspond to the pseudogap energy. Two observations suggest the latter scenario to be more applicable to the underdoped cuprates. First, we find no evidence for quantum oscillations originating from the hole pocket, either by direct observation or by way of magnetic breakdown combination frequencies. Second, the Fermi arc, which refers to the region in momentum space over which the photoemission spectral weight is strongest, is seen to be very similar in length to the sides of the electron pocket in Figs.~\ref{schematic}a and ~\ref{schematic}b (for both Hg1201 and Y123). The spectral weight drops off precipitously beyond the tips of the pocket. We note that while low frequency QOs in Y123 have been attributed to small hole pockets, this low frequency could also originate from Stark quantum interference effects associated with bilayer splitting~\cite{sebastian14}. Alternatively, the pseudogap phenomena could also menifest as strong scattering at the antinodal regions, thus preventing an observation of such pockets in Hg1201. Recent high-temperature normal-state transport measurements in Hg1201 have also been interpreted in terms of Fermi-liquid-like~\cite{mirzaei13,chan14} Fermi arcs~\cite{barisic13}.

Our findings in Hg1201 have direct implications for the interpretation of QO measurements made in other cuprate materials. If we assume a similar gap size between Hg1201 and Y123, the large $E_{\rm g}$ suggests that magnetic breakdown combination frequencies involving the electron and small hole pocket~\cite{doiron15,allais14} cannot be responsible for the complicated beat pattern associated with closely spaced frequencies in Y123~\cite{audouard09,sebastian14} and Y124~\cite{tan15}. The splitting of the main frequency into two or more components must therefore be the consequence of the bilayer coupling in those systems~\cite{sebastian14,tan15,briffa15}, or a stronger interlayer $c$-axis hopping. 

The biaxial reconstruction confirmed here for Hg1201 has also been proposed for Y123\cite{sebastian14, harrison15, briffa15}, which is supported by ultrasound measurements in high-fields~\cite{leboeuf13}. However, x-ray measurements show apparent local-stripe CDW domains at high temperatures~\cite{comin15}, which presumably become long-ranged and possibly arranged in a criss-cross pattern of stripes at low temperatures and high-fields~\cite{maharaj14,maharaj15}. Recent X-ray measurements on Y123 show a new magnetic field induced three-dimensional CDW centered at $c$-axis wave-vector $L=1$~r.l.u. ~\cite{gerber15} only along the CuO chain directions~\cite{chang16} which breaks the mirror symmetry of the CuO$_2$ bilayers. The role of bilayer coupling and CuO chains for this stripe-like ordering tendency is still an open question, and its relevance for Hg1201 which features neither is unclear. The stripe picture is attractive because of natural analogies to single-layered La-based cuprates~\cite{tranquada95} and its implications for the role of nematicity (broken planar rotational symmetry) for the cuprate phase diagram~\cite{kivelson03}. However, neutron scattering experiments have found that the typical signatures of spin stripes are absent in the  magnetic excitations of Hg1201~\cite{chan16}. Despite the appearance of a new uniaxial 3D order in Y123, the CDW wavevector, with a smaller $c$-axes correlation length, is still clearly observed in both planar directions in magnetic fields up to $\sim 17$~ T~\cite{chang16}. For Hg1201, we have shown here that the CDW that causes the Fermi-surface reconstruction is biaxial. It thus remains an open question as to whether electronic nematicity is generic to the cuprates, particularly in tetragonal Hg1201.

\section{Methods}

{\bf Samples.} Hg1201 single crystals were grown using a self-flux method~\cite{zhao06}. As grown crystals have $T_{\rm c}\approx 80$~K. Post-growth heat treatment in N$_2$ atmosphere at 400$^o$C and 450$^o$C was used to achieve $T_{\rm c} = 74$~K (hole concentration $p=0.097$) and $T_{\rm c} = 71(2)$~K ($p=0.09$) respectively. $T_{\rm c}$ was determined from DC susceptibility measurements.The 95$\%$ level transition width of both samples is $2$~K. The hole concentration $p$ is determined based on the phenomenological Seebeck coefficient scale \cite{yamamoto00}.

The YBCO crystal was flux grown and heat treated to obtain oxygen content $x=0.58$ with $T_{\rm c}=60$~K and hole doping $p=0.106$ at the University of British Columbia, Canada~\cite{liang06}.

{\bf Pulsed field measurements.} High magnetic field measurements were performed at the Pulsed-Field Facility at Los Alamos National Laboratory. The magnet system used consists of an inner and outer magnet. The outer magnet is first generator driven relatively slowly ($\sim3$~s total width) between 0 T and 37 T, followed by a faster ($\sim 15$~ms) capacitor bank driven pulse to $90$~T.   

{\bf Fitting quantum oscillations.} We fit the field dependence to $\Delta f/f=(a_0+a_1B+a_2B^2+ \dots)(1+A_{\rm osc})$, where the first term is a polynomial representing the non-oscillatory background and $A_{\rm osc}$ is the oscillatory component. In the case of a single Fermi surface cylinder, the QOs are described by the Lifshitz-Kosevitch form~\cite{shoenberg}
\begin{equation}
 A_{\rm osc}=A_0 R_T R_{\rm D} R_{\rm S}R_{\rm MB} R_{\rm W}\cos{\Big[}2\pi{\Big(}\frac{F}{B}-\gamma{\Big)}{\Big]}{\rm ,}
\label{lk1}\end{equation} 
where $F$ is the frequency of QOs, $\gamma$ is the phase and $A_0$ is a temperature and field independent pre-factor. Here, $R_T$, $R_{\rm D}$, $R_{\rm S}$, $R_{\rm MB}$ and $R_{\rm W}$ are the thermal, Dingle, spin, magnetic breakdown and warping damping factors, respectively~\cite{sebastian14,ramshaw11}. $R_T={\alpha T/[B}{\sinh(\alpha T/B)]}$ where $\alpha ={2\pi^2k_Bm^*}/{(e\hbar)}$ accounts for the thermal broadening of the Fermi-Dirac distribution relative to the cyclotron energy and $m^\ast=$~2.7~$m_{\rm e}$, determined for one of our samples as shown in Fig.~\ref{Tdep}, is the quasiparticle effective mass ($m_{\rm e}$ being the free electron mass). Meanwhile, $R_{\rm D}={\rm exp}(-\pi l_{\rm c}/l)$, where $l_{\rm c}=\sqrt{2\hbar F/e}/B$ is the cyclotron radius and $l$ is the mean free path. To lowest order, warping of a cylindrical Fermi surface leads to an amplitude reduction factor of the form $R_{\rm w}={\rm J}_0(2\pi{\Delta F_c}/{B})$ in which ${\rm J}_0$ is a zeroth order Bessel function and $2\Delta F_{\rm c}\approx 4t_\perp m^\ast/{(\hbar e)}$ is the difference in frequency between the minimum and maximum cross-sections of the warped cylinder. Since our experiments are performed at fixed angle (i.e. ${\bf B}\parallel c$), we neglect $R_{\rm S}$ by setting it to unity. As discussed in the main text, our data shows no signatures of magnetic breakdown tunneling or warping, thus we also set $R_{\rm MB}$ and $R_{\rm W}$ to unity. Limits on these two terms are discussed in the Results section.

{\bf Calculation of reconstructed Fermi surface.}
The unreconstructed Fermi surface is calculated with the dispersion $\epsilon_{\bf k} = -2t(\phi_x+\phi_y)-4t^\prime\phi_x\phi_y-2t^{\prime\prime}(\phi_{2x}+\phi_{2y})-4t^{\prime\prime}(\phi_{2x}\phi_y+\phi_{2y\phi_x})-\mu$ where the tight-binding parameters are $(t,t^{\prime},t^{\prime\prime},t^{\prime\prime\prime}) = (0.46,-0.105,0.08,-0.02)~{\rm eV}$~\cite{das12} for Hg1201 and $(0.35,-0.112,0.007,0)~{\rm eV}$ for YBCO. $\mu$ is the chemical potential and $\phi_{nx}=\cos(nk_x)$ and $\phi_{ny}=\cos(nk_y)$ where $k_x$ and $k_y$ are the planar wavevectors. We required that the tight-binding parameters produce a Fermi surface in agreement with the photoemission data and have carrier number $1+p$ where $p=0.12$ and $0.11$ for the Hg1201 and Y123 samples on which the photoemission data were taken. Hence, $1+p = 2A_{\rm UFS}/A_{\rm UBZ}$, where $A_{\rm UFS}$ and $A_{\rm UBZ}$ are the areas of the unreconstructed Fermi surface and Brillouin zone respectively. Before calculating the reconstructed Fermi surface, only $\mu$ is adjusted to match the hole doping $p=0.095$ and $p=0.106$ on which the QO data was taken for Hg1201 and Y123 respectively.

Following Ref.~\cite{harrison11}, the reconstructed Fermi surface is determined by diagonalizing a Hamiltonian considering translations of the biaxial CDW wavevector  ${\bf k}\rightarrow {\bf k}+n_xQ_{\rm CDW}~{\bf \hat{x}}+n_yQ_{\rm CDW}~{\bf \hat{y}} $, where $n_x$ and $n_y$ are the number of translations in the planar directions. Strictly speaking, reconstruction by observed incommensurate CDW wavevectors requires an infinite number of terms in the Hamiltonian to obtain all the bands. However, since $\Delta\ll t$, the inclusion of high order terms in the Hamiltonian gives rise to a hierarchy of higher order gaps that are exponentially small, and thus do not effect the primary closed orbits resulting from our calculation, which we restrict to nine terms. Supplementary Figure~3 shows all the bands resulting from our reconstruction calculation. 

We use $\Delta_{\rm CDW}/t=$~0.1 for the ratio of the CDW order parameter magnitude to the in-plane hopping~\cite{allais14}.  This implies $\Delta_{\rm CDW} = 46$~meV, based on band structure determination of $t$ \cite{das12}, which is larger than the lower bound value determined from our analysis of magnetic breakdown tunneling in the main text, but sufficiently small that it does not adversely affect the sizes of the pockets. Reducing the ratio to zero increases the area of the reconstructed pockets by only $\approx$~3~\%.

{\bf Acknowledgments}\\
This work, performed at Los Alamos National Lab, was supported by the US Department of Energy BES ``Science at 100 T" grant no. LANLF100. The National High Magnetic Field Laboratory - PFF facility is funded by the National Science Foundation Cooperative Agreement No. DMR-1157490, the State of Florida, and the U.S. Department of Energy. Work at the University of Minnesota was supported by the Department of Energy, Office of Basic Energy Sciences, under Award No. DE-SC0006858. N.B. acknowledges the support of FWF project P2798. We thank Ruixing Liang, W. N. Hardy and D. A. Bonn at UBC, Canada for generously supplying the Y123 crystal measured as part of this work. We aknowledge fruitful discussion with S.E. Sebastian. We also thank the Pulsed Field Facility, Los Alamos National Lab engineering and technical staff for experimental assistance.

{\bf Correspondence}\\
Correspondence should be addressed to M.K.C. (mkchan@lanl.gov) and N.H. (nharrison@lanl.gov).

{\bf Author Contributions}\\
M.K.C., N.H., R.D.M., B.J.R.  K.A.M. and N.B. designed and performed contactless resistivity measurements in pulsed magnetic fields on Hg1201 and Y123. M.K.C. synthesized and prepared the Hg1201 samples. N.H. supervised the work at Los Alamos National Lab. M.G. supervised the work at the University of Minnesota. M.K.C. and N.H. wrote the manuscript with critical input from all authors.

\pagebreak

\bibliography{mercurywave}

\end{document}